\documentclass[twocolumn]{aastex6}
\usepackage{graphicx}
\usepackage{amssymb,amsfonts,amsmath,amstext,amsgen,amsopn,amsxtra,indentfirst,times}

\begin{document}

\title{Analytical Models of Exoplanetary Atmospheres. IV. Improved Two-stream Radiative Transfer for the Treatment of Aerosols}

\author{Kevin Heng\altaffilmark{1,2,3}}
\author{Daniel Kitzmann\altaffilmark{1}}
\altaffiltext{1}{University of Bern, Center for Space and Habitability, Gesellschaftsstrasse 6, CH-3012, Bern, Switzerland.  Emails: kevin.heng@csh.unibe.ch, daniel.kitzmann@csh.unibe.ch}
\altaffiltext{2}{Johns Hopkins University, Department of Earth and Planetary Sciences, 301 Olin Hall, Baltimore, MD 21218, U.S.A.}
\altaffiltext{3}{Johns Hopkins University, Department of Physics and Astronomy, Bloomberg Center for Physics and Astronomy, Baltimore, MD 21218, U.S.A.}

\begin{abstract}
We present a novel generalization of the two-stream method of radiative transfer, which allows for the accurate treatment of radiative transfer in the presence of strong infrared scattering by aerosols.  We prove that this generalization involves only a simple modification of the coupling coefficients and transmission functions in the hemispheric two-stream method.  This modification originates from allowing the ratio of the first Eddington coefficients to depart from unity.  At the heart of the method is the fact that this ratio may be computed once and for all over the entire range of values of the single-scattering albedo and scattering asymmetry factor.  We benchmark our improved two-stream method by calculating the fraction of flux reflected by a single atmospheric layer (the reflectivity) and comparing these calculations to those performed using a 32-stream discrete-ordinates method.  We further compare our improved two-stream method to the two-stream source function (16 streams) and delta-Eddington methods, demonstrating that it is often more accurate at the order-of-magnitude level.  Finally, we illustrate its accuracy using a toy model of the early Martian atmosphere hosting a cloud layer composed of carbon-dioxide ice particles.  The simplicity of implementation and accuracy of our improved two-stream method renders it suitable for implementation in three-dimensional general circulation models.  In other words, our improved two-stream method has the ease of implementation of a standard two-stream method, but the accuracy of a 32-stream method.
\end{abstract}


\section{Introduction}
\label{sect:intro}

Two-stream solutions have been studied for decades in the context of atmospheres and come in various flavors \citep{s05,chandra60,mihalas70,mihalas78,mw80,gy89,toon89,pierrehumbert10,heng17}.  They originate from a neat mathematical trick: instead of solving the radiative transfer equation for the intensity, one solves for its moments.  Besides the loss of angular information, the two-stream solution performs poorly when aerosols reside in the model atmosphere.  A longstanding result, based on geomorphic evidence, that early Mars was able to harbor liquid water on its surface (see \citealt{wordsworth16} for a review), due to the scattering greenhouse effect (e.g., \citealt{pierrehumbert10,hhps12}) mediated by carbon-dioxide ice clouds \citep{fp97}, was called into question because the original two-stream calculation over-estimated the degree of warming \citep{kitzmann16}.  Mars teaches us the lesson that the choice of radiative transfer method may alter the qualitative conclusion of a study, and inspires us to improve the accuracy of the two-stream solution in order to apply it broadly to exoplanetary atmospheres.

\begin{figure}
\begin{center}
\includegraphics[width=\columnwidth]{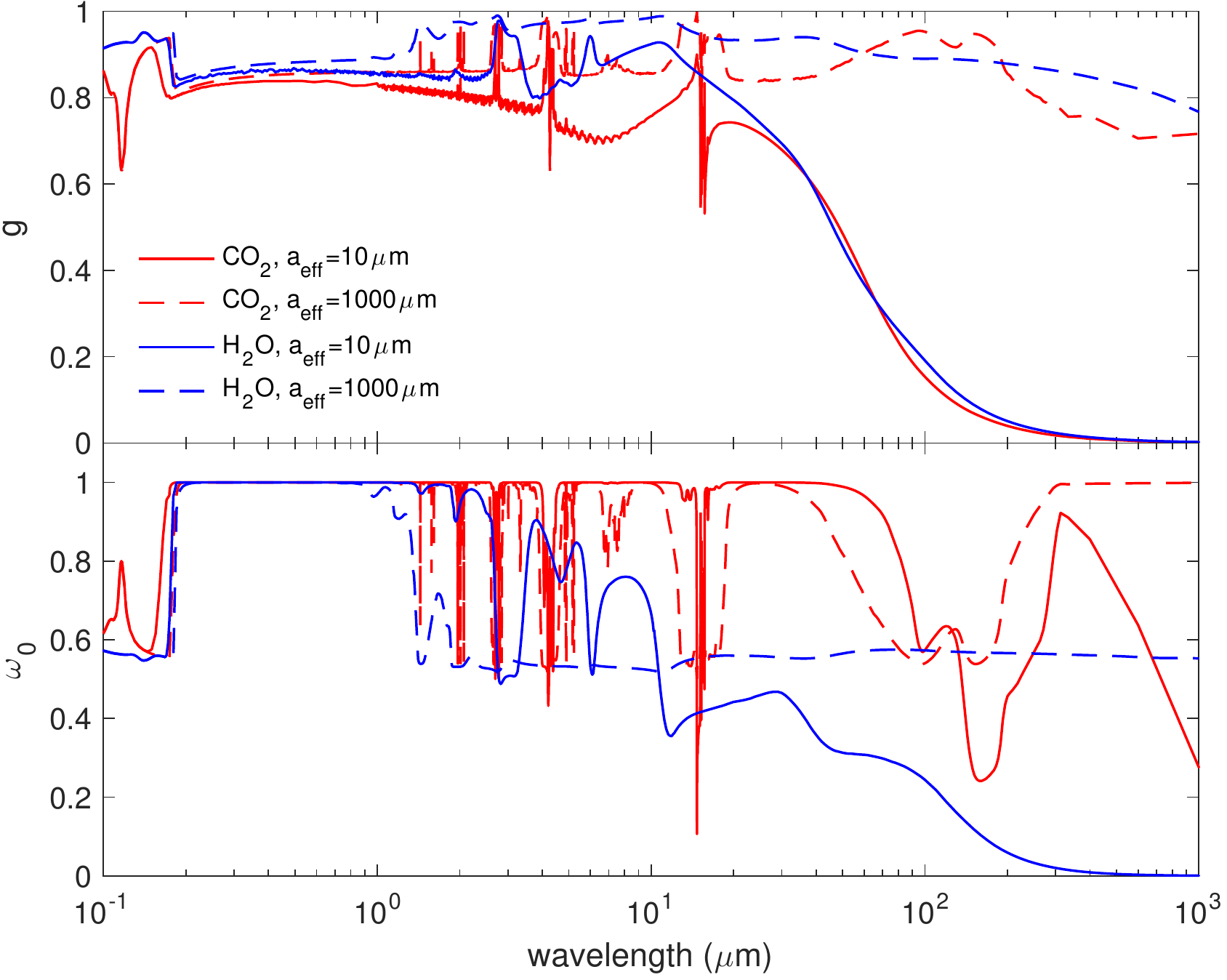}
\end{center}
\vspace{-0.1in}
\caption{Scattering asymmetry factor (top panel) and single-scattering albedo (bottom panel) of water ice versus carbon-dioxide ice.  The size distribution of particles follows a gamma distribution as stated in equation (1) of \cite{kitzmann13}, where the effective radius is $a_{\rm eff}$.}
\label{fig:properties}
\end{figure}

The main source of error appears to be the over-estimation of the amount of infrared radiation reflected by aerosols, which leads to an over-estimation of the scattering greenhouse effect.  On Earth, this effect is subdued because water clouds are strong infrared absorbers but weak infrared scatterers \citep{pierrehumbert10}.  On Mars, it is pronounced because carbon-dioxide ice clouds scatter infrared radiation strongly \citep{kitzmann13}.  Figure \ref{fig:properties} illustrates these differences.  In general, we expect the two-stream method to perform poorly in the presence of medium-sized to large aerosols that have single-scattering albedos between 0.5 and 1 in the infrared range of wavelengths.  This shortcoming motivates us to design an improved two-stream method that calculates the amount of reflected radiation accurately.  Operationally, we accomplish this feat by revisiting the formalism surrounding the Eddington coefficients previously elucidated by \cite{hml14}.  

\begin{figure*}
\begin{center}
\vspace{-0.2in}
\includegraphics[width=0.7\columnwidth]{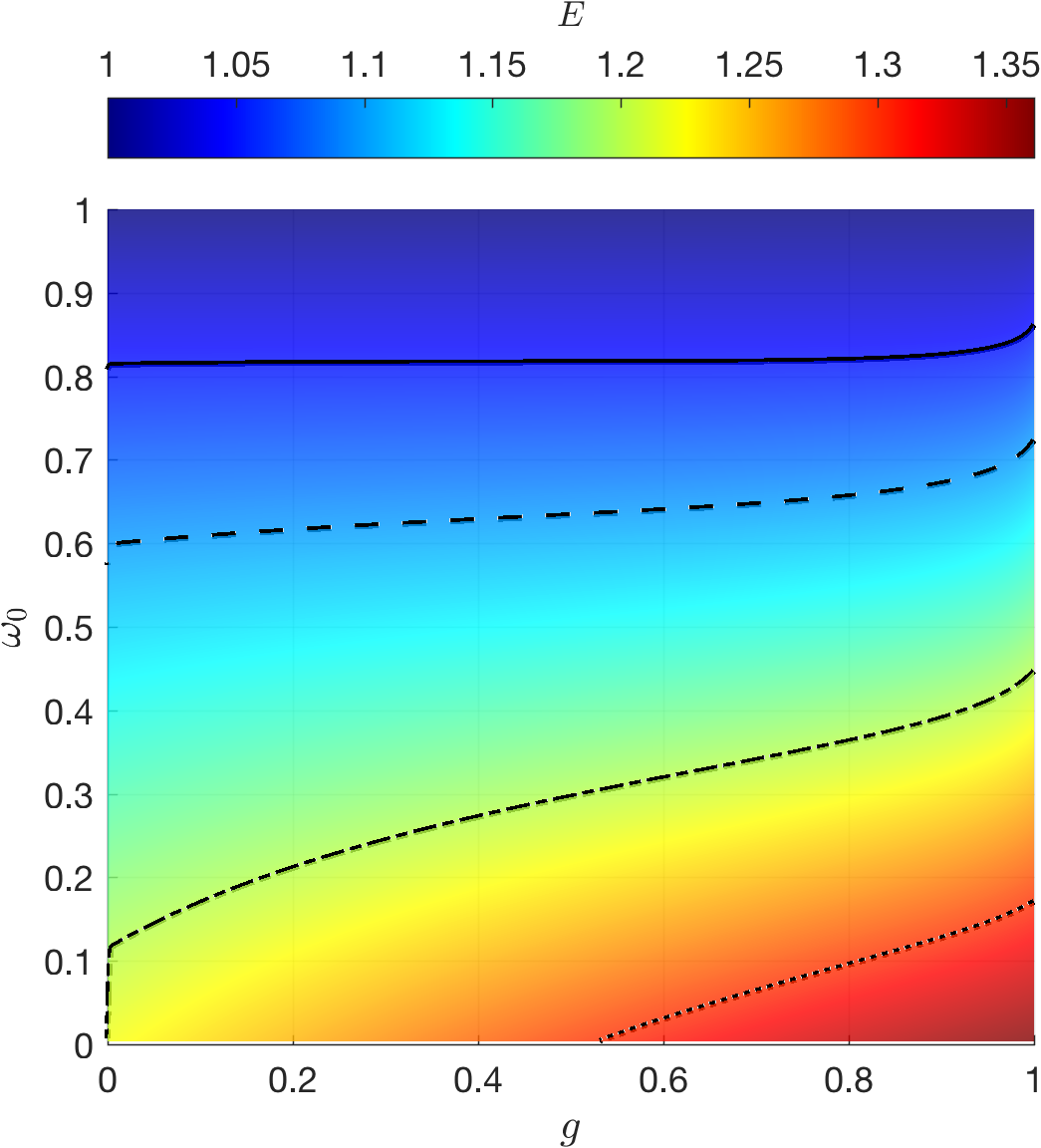}
\includegraphics[width=\columnwidth]{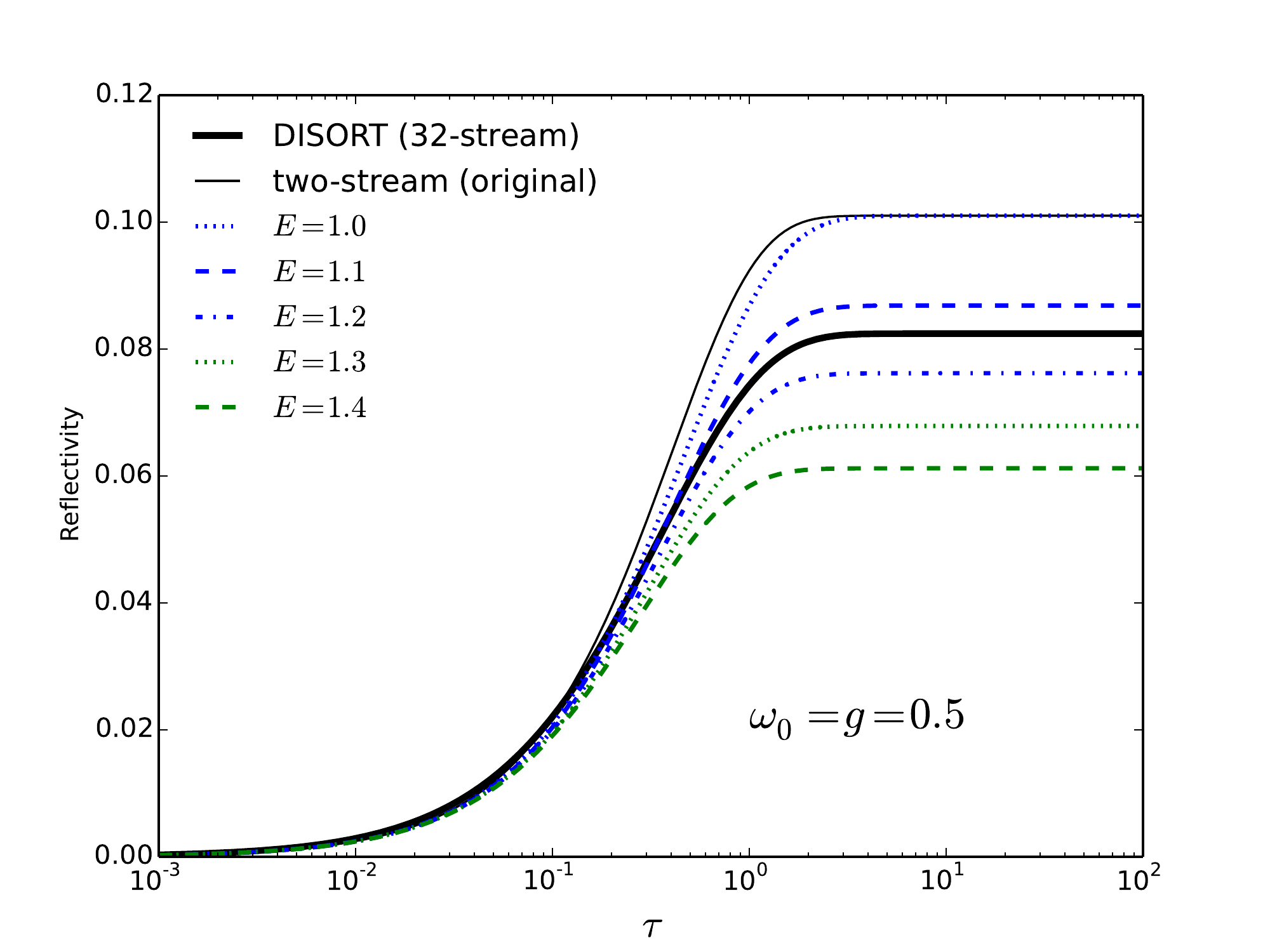}
\includegraphics[width=\columnwidth]{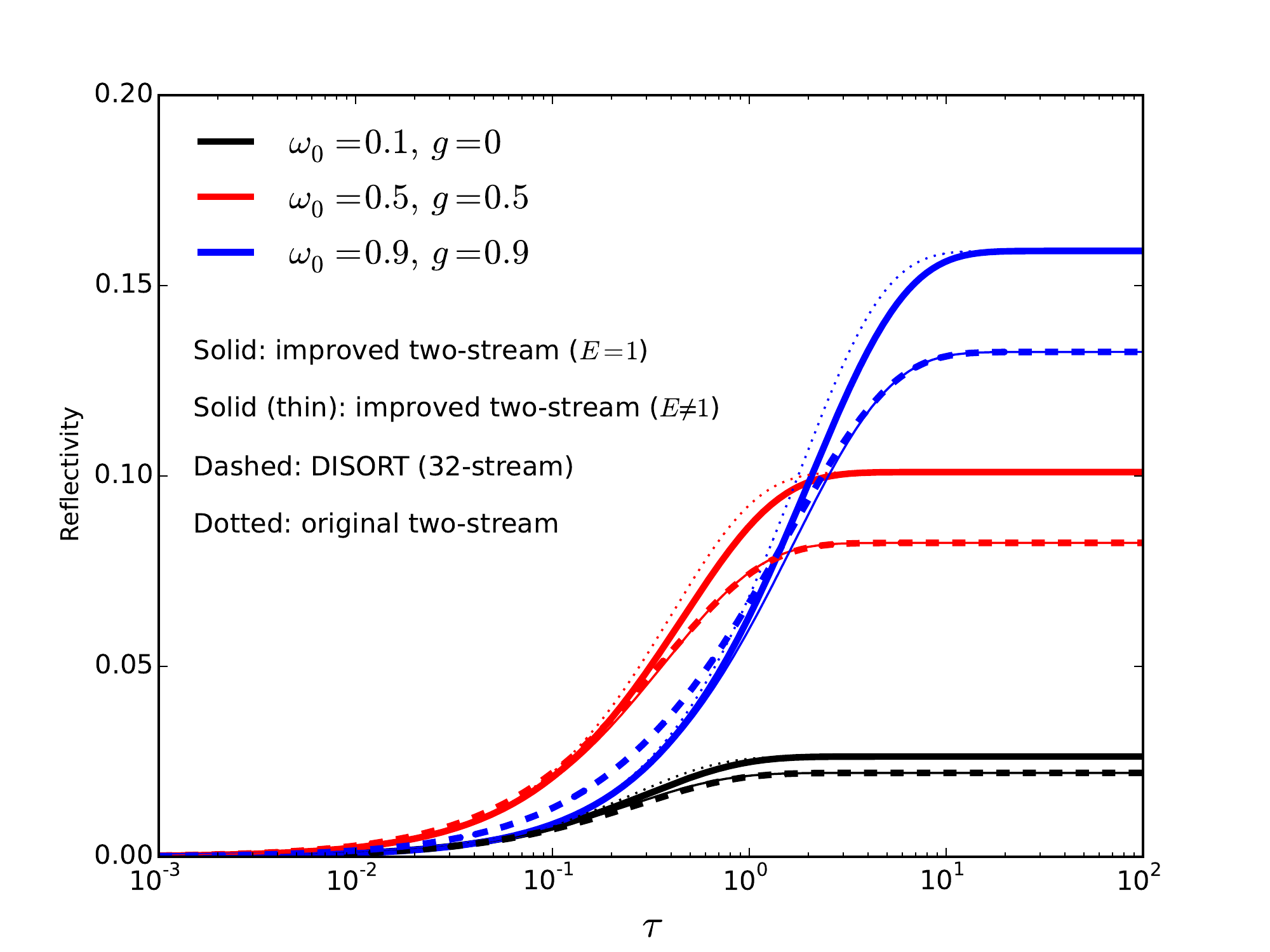}
\includegraphics[width=\columnwidth]{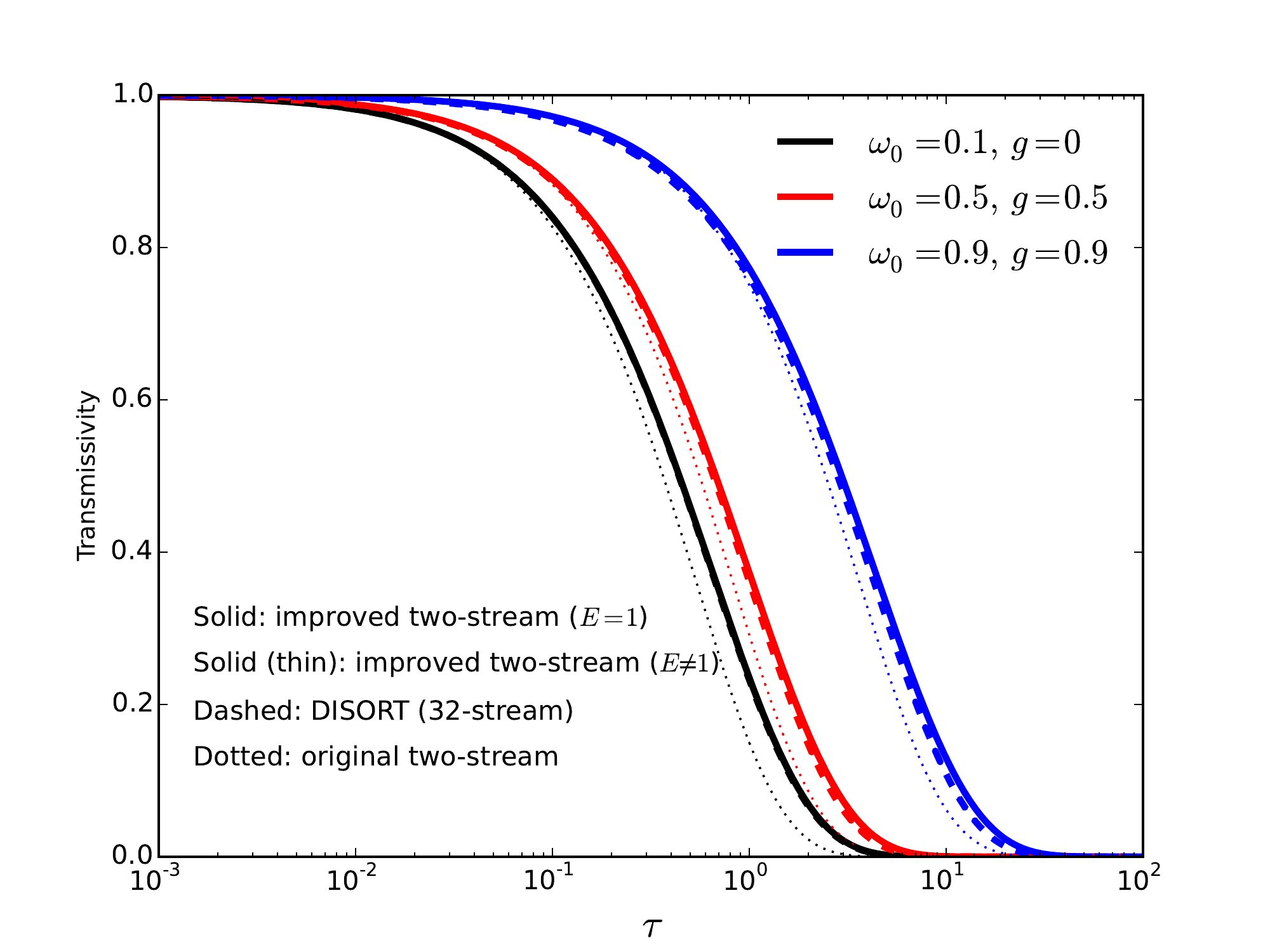}
\end{center}
\vspace{-0.2in}
\caption{Top-left panel: Ratio of Eddington coefficients, $E$, as a function of single-scattering albedo ($\omega_0$) and scattering asymmetry factor ($g$).  The original, hemispheric two-stream method always has $E=1$.  The solid, dashed, dot-dashed and dotted curves are for $E=1.05, 1.1, 1.2$ and 1.3, respectively.  Top-right panel: Reflectivity for a medium-sized aerosol computed using the original, hemispheric two-stream method, our improved two-stream method (with various values of $E$) and a 32-stream discrete ordinates method (via the \texttt{DISORT} code) versus the optical depth of the atmospheric layer ($\tau$).  Bottom-left panel: Reflectivity for small, medium-sized and large aerosols.  Bottom-right panel: Transmissivity for small, medium-sized and large aerosols, where the pair of solid curves completely overlap.}
\label{fig:efactor}
\end{figure*}

Specifically, we relax the assumption that the first Eddington coefficients\footnote{As already noted in \cite{hml14}, there is no consensus on how to number/order these Eddington coefficients, and we use the convention of \cite{hml14}.} are equal and allow their ratio to depart from unity.  The top-left panel of Figure \ref{fig:efactor} shows our calculations for this ratio, $E$.  We also show calculations for single atmospheric layers populated by aerosols with fixed values of the single-scattering albedo ($\omega_0$) and scattering asymmetry factor ($g$).  We consider single atmospheric layers, because if one attains understanding (and accuracy) for a single layer, then it straightforwardly generalizes to an arbitrary number of layers in a model atmosphere.  For the sake of discussion, we refer to small, medium-sized and large aerosols as having $\omega_0=0.1$ and $g=0$ (isotropic scattering), $\omega_0=g=0.5$ and $\omega_0=g=0.9$ (predominantly forward scattering), respectively.  We will explore other choices later.  To simplify terminology, we term the fraction of flux reflected and transmitted by an atmospheric layer the ``reflectivity" and ``transmissivity", respectively.  In the example of a layer populated by medium-sized aerosols (top-right panel of Figure \ref{fig:efactor}), we see that the original, hemispheric two-stream solution (e.g., \citealt{pierrehumbert10,hml14}) over-estimates the true solution, which is computed using a 32-stream discrete-ordinates method via the open-source \texttt{DISORT} computer code \citep{s88,hamre13}.  Our improved two-stream solution with $E=1$ matches the reflectivity computed by the hemispheric two-stream method well; deviations are due to modifications we have made to the transmission function, as we will discuss.  As $E$ is varied from 1 to 1.4, we see that the reflectivity varies rather sensitively.  The true solution is matched by a value of $E$ between 1.1 and 1.2.  This example illustrates that small variations of $E$ from unity allow us to improve the accuracy of the two-stream solution drastically.

Following through on this property of $E$, the bottom-left panel of Figure \ref{fig:efactor} shows calculations of the reflectivity for atmospheric layers with small, medium-sized and large aerosols.  For each calculation, the value of $E$ has been chosen to match the reflectivity \textit{by construction}; in \S\ref{sect:formalism}, we will explain in detail how this is accomplished.  For all three calculations, the original, hemispheric two-stream method over-estimates the reflectivity by $\sim 10\%$.  For completeness, we also show the transmissivity associated with these three examples (bottom-right panel of Figure \ref{fig:efactor}), where we see that the discrepancies between the hemispheric two-stream calculations and the true solutions are less pronounced.

The overaching goal of the present study is to elucidate the theory behind the improvement of the two-stream method and the calculation of $E$ (presented in \S\ref{sect:formalism}).  We further demonstrate that our improved two-stream method rivals or betters the two-stream source function method of \cite{toon89}, which is widely implemented in the exo-atmospheres literature, in both accuracy and simplicity of implementation (in \S\ref{sect:results}).  We discuss the implications of our findings in \S\ref{sect:discussion}.  The present study is the fourth in a series of papers devoted to constructing analytical models for exoplanetary atmospheres to both aid in the development of intuition and provide algorithms for computation, following \cite{hw14} (for shallow-water fluid dynamics), \cite{hml14} (for two-stream radiative transfer) and \cite{ht16} (for equilibrium chemistry).

\section{Generalizing the two-stream formalism}
\label{sect:formalism}

In the two-stream formalism, the reflectivity and transmissivity of a single layer are respectively \citep{hml14}
\begin{equation}
f_{\rm T} = \frac{\left( \zeta_-^2 - \zeta_+^2 \right) {\cal T} }{\left( \zeta_- {\cal T} \right)^2 - \zeta_+^2}, ~f_{\rm R} = \frac{\zeta_- \zeta_+ \left( 1 - {\cal T}^2 \right) }{\zeta_+^2 - \left( \zeta_- {\cal T} \right)^2},
\end{equation} 
where $\zeta_\pm$ are the coupling coefficients and ${\cal T}$ is the transmission function.  The coupling coefficients relate the relative strength of transmission versus reflection, and generally depend on $\omega_0$ and $g$.  When the layer is transparent, we have ${\cal T}=1$, $f_{\rm T}=1$ and $f_{\rm R}=0$.  When the layer is opaque, we have ${\cal T}=0$, $f_{\rm T}=0$ and $f_{\rm R} = \zeta_-/\zeta_+$.  These asymptotic limits suggest that plausible improvements to the two-stream solution are accomplished by modifying the coupling coefficients.

In the present study, we focus on the reflectivity and transmissivity of a single layer, and not its emissivity (blackbody emission), because previous studies dealing with aerosols have shown that the largest sources of error originate from the reflectivity \citep{kitzmann13,kitzmann16}.  It has been previously shown that the expressions for the coupling coefficients with the hemispheric closure conserve energy by construction \citep{hml14}.

To modify the coupling coefficients, we need to understand their physical origin.  In sacrificing accuracy for simplicity, the two-stream solutions contain an ambiguity: ratios of the moments of the intensity (mean intensity, flux, radiation pressure) are assumed to be constants known as Eddington coefficients \citep{mw80,toon89,pierrehumbert10,hml14,heng17}.  Two of these Eddington coefficients are set to be equal by enforcing the condition of radiative equilibrium in the limit of pure scattering \citep{toon89}.  This occurs when the single-scattering albedo is exactly unity.  However, the two-stream solutions formally break down at exactly $\omega_0=1$ \citep{toon89,hml14} and this limit is rarely reached in practice, which render this condition academic.  If we relax this condition, then the coupling coefficients have a more general form,
\begin{equation}
\zeta_\pm \equiv \frac{1}{2} \left[ 1 \pm \sqrt{\frac{E - \omega_0}{E - \omega_0 g}} \right].
\label{eq:zetas}
\end{equation}
where $E$ is the ratio of the Eddington coefficients.  In the original two-stream solutions with the hemispheric (or hemi-isotropic) closure, we have $E=1$.  The first improvement is to use
\begin{equation}
E = \frac{\omega_0 \left( 1 - g r^2 \right)}{1 - r^2}
\label{eq:ee}
\end{equation}
to compute $E\ne1$ values for use in the coupling coefficients.  Here, we have $r \equiv (1-R_\infty)/(1+R_\infty)$ and $R_\infty$ is the asymptotic value of the reflectivity when a layer is opaque ($\tau \gg 1$).  The preceding expression is derived by setting $R_\infty = \zeta_-/\zeta_+$ and using equation (\ref{eq:zetas}).  It is worth emphasizing that this improvement ensures the asymptotic reflectivity matches the true solution \textit{by construction,} as long as we have a way of computing $R_\infty$.

In Figure \ref{fig:efactor} (top-left panel), the grid of values for $E$ obtains from computing $R_\infty$ using the \texttt{DISORT} code \citep{s88,hamre13}, which uses the discrete-ordinates method of radiative transfer \citep{chandra60}.  We use 32-stream \texttt{DISORT} calculations as the ground truth.  It should be emphasized that this is the \textit{entire} parameter space of interest for aerosols embedded in atmospheres.  Part of the simplicity of the method is that the function $E(\omega_0, g)$ only needs to be computed once.  It may then be stored and used for all future calculations.

The second improvement we make is to modify the transmission function.  In the limit of pure absorption, the exact solution of the radiative transfer equation yields ${\cal T} = 2 E_3(\tau)$, an expression that formally integrates over all angles \citep{hml14}.  Here, $E_3$ is the exponential integral of the third order \citep{arfken} and $\tau$ is the optical depth of the atmospheric layer.  For transmission, we use ${\cal T} = 2 E_3 \left( \tau^\prime \right)$, where $\tau^\prime = \tau \sqrt{(1-\omega_0)(1-\omega_0 g)}$ and the additional factor derives from the hemispheric two-stream solution \citep{hml14}.  For reflection, we use ${\cal T} = 2 E_3\left( \tau^{\prime \prime} \right)$ and $\tau^{\prime \prime} = \tau \sqrt{(E-\omega_0)(E-\omega_0 g)}$.  

\section{Results}
\label{sect:results}

\subsection{Comparison to \cite{toon89}}

\begin{figure*}
\begin{center}
\vspace{-0.2in}
\includegraphics[width=\columnwidth]{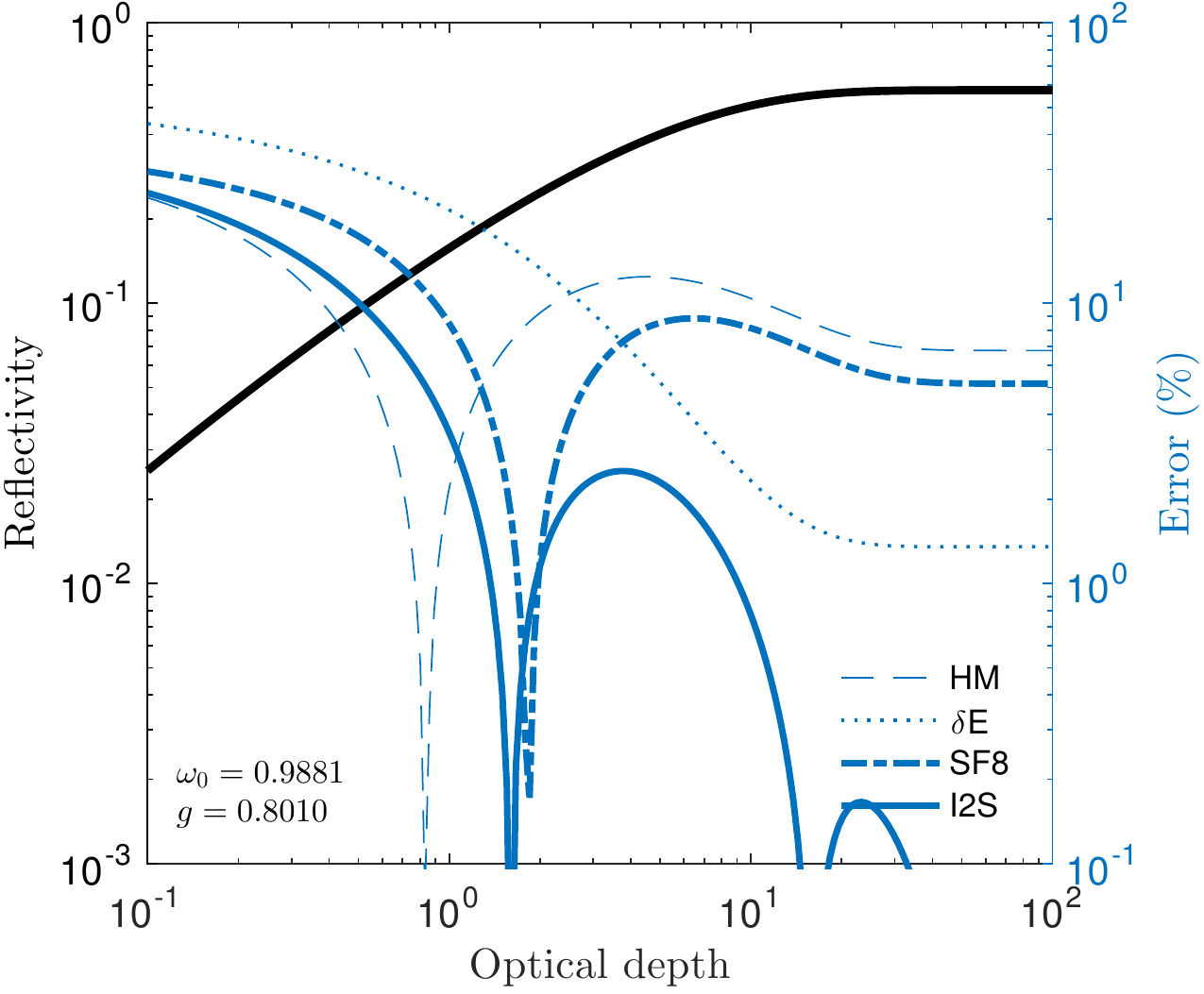}
\includegraphics[width=\columnwidth]{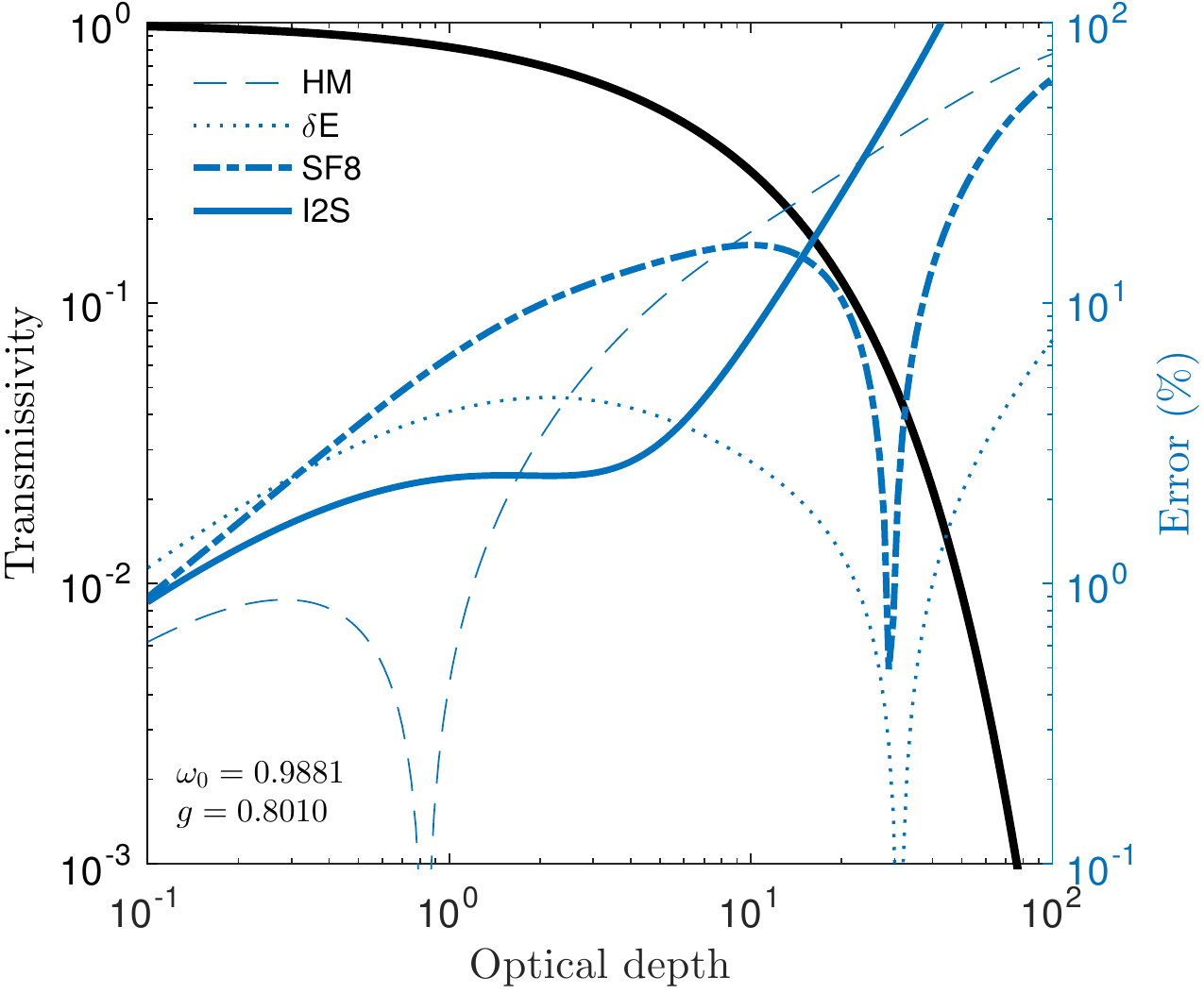}
\includegraphics[width=\columnwidth]{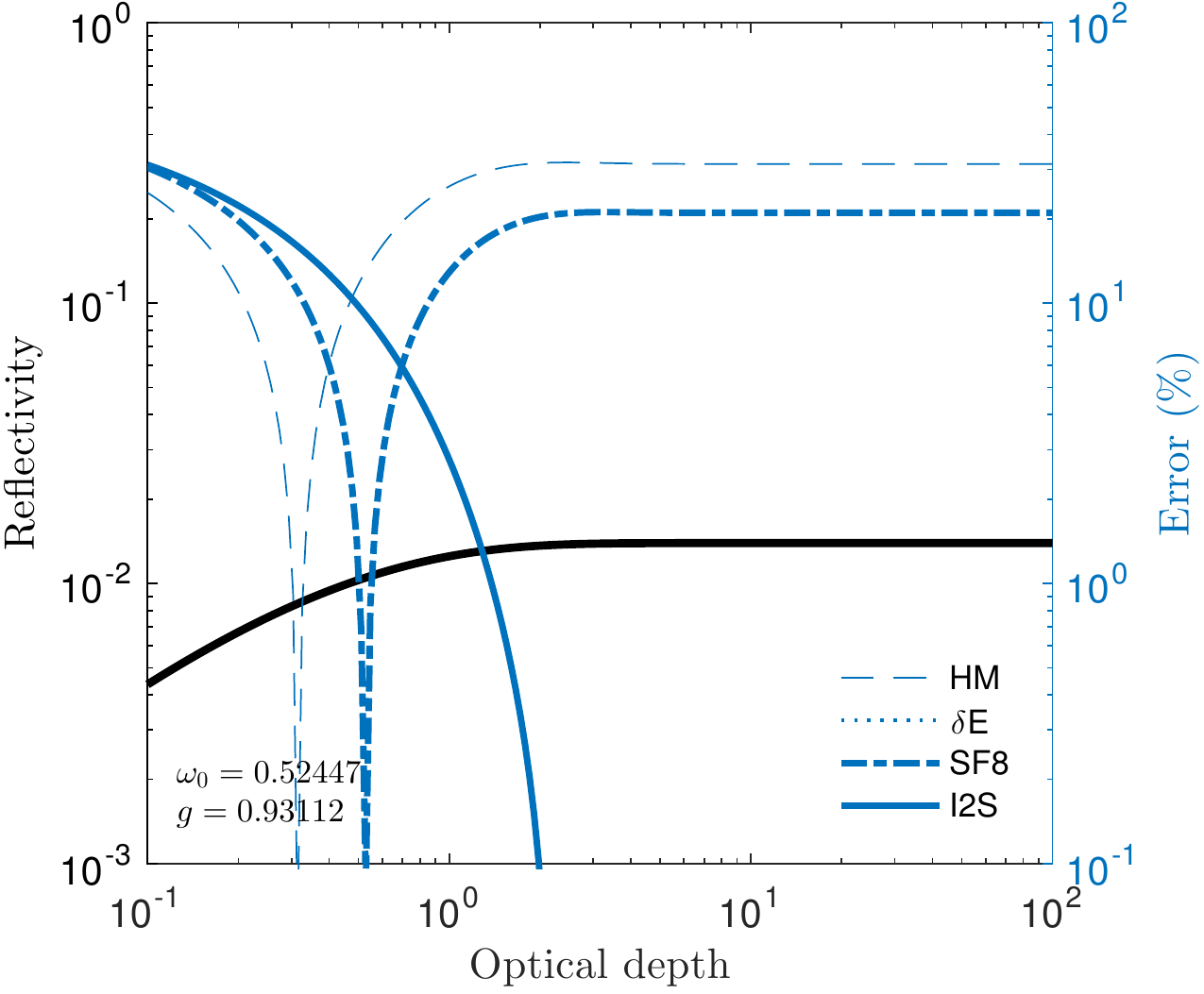}
\includegraphics[width=\columnwidth]{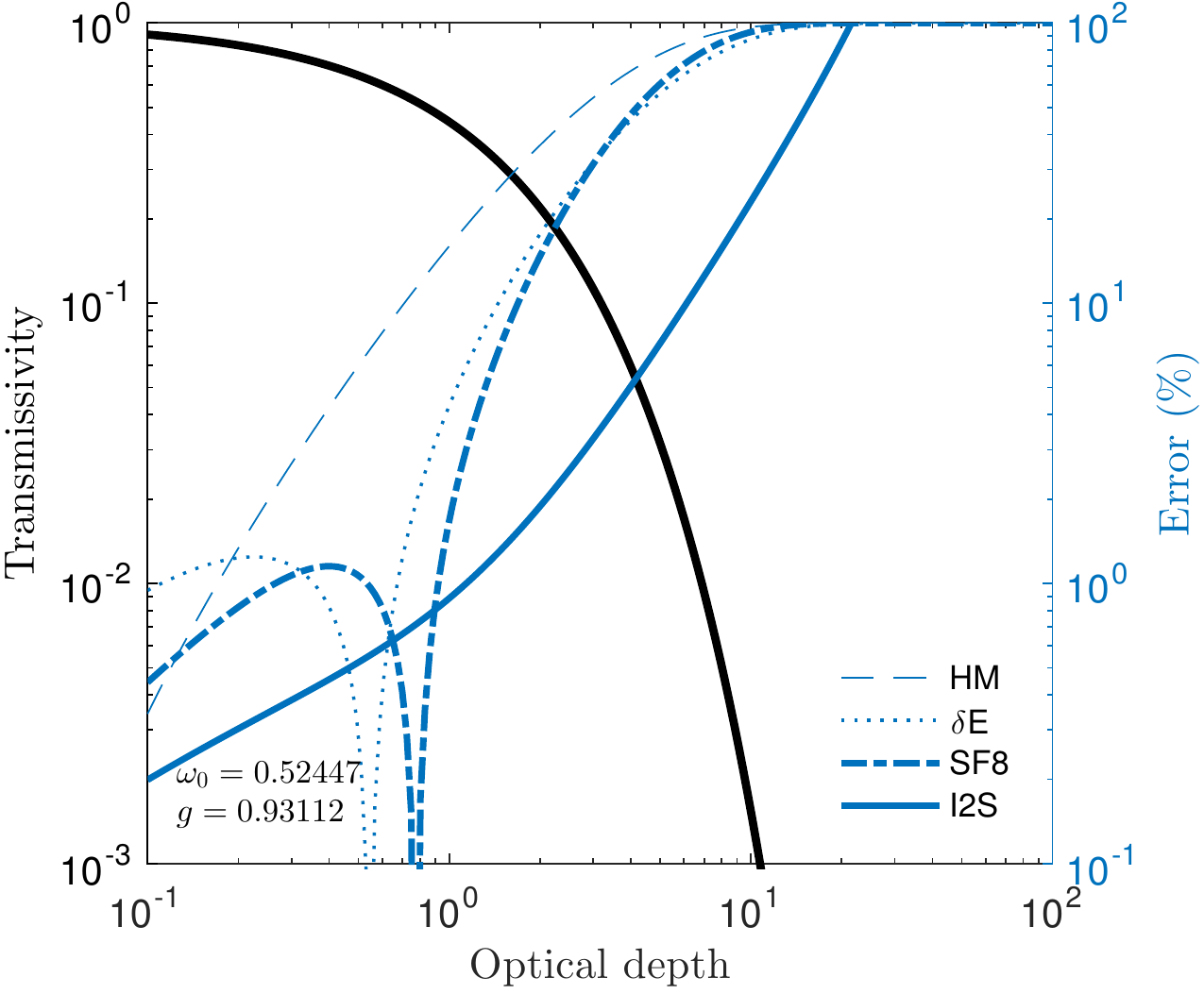}
\includegraphics[width=\columnwidth]{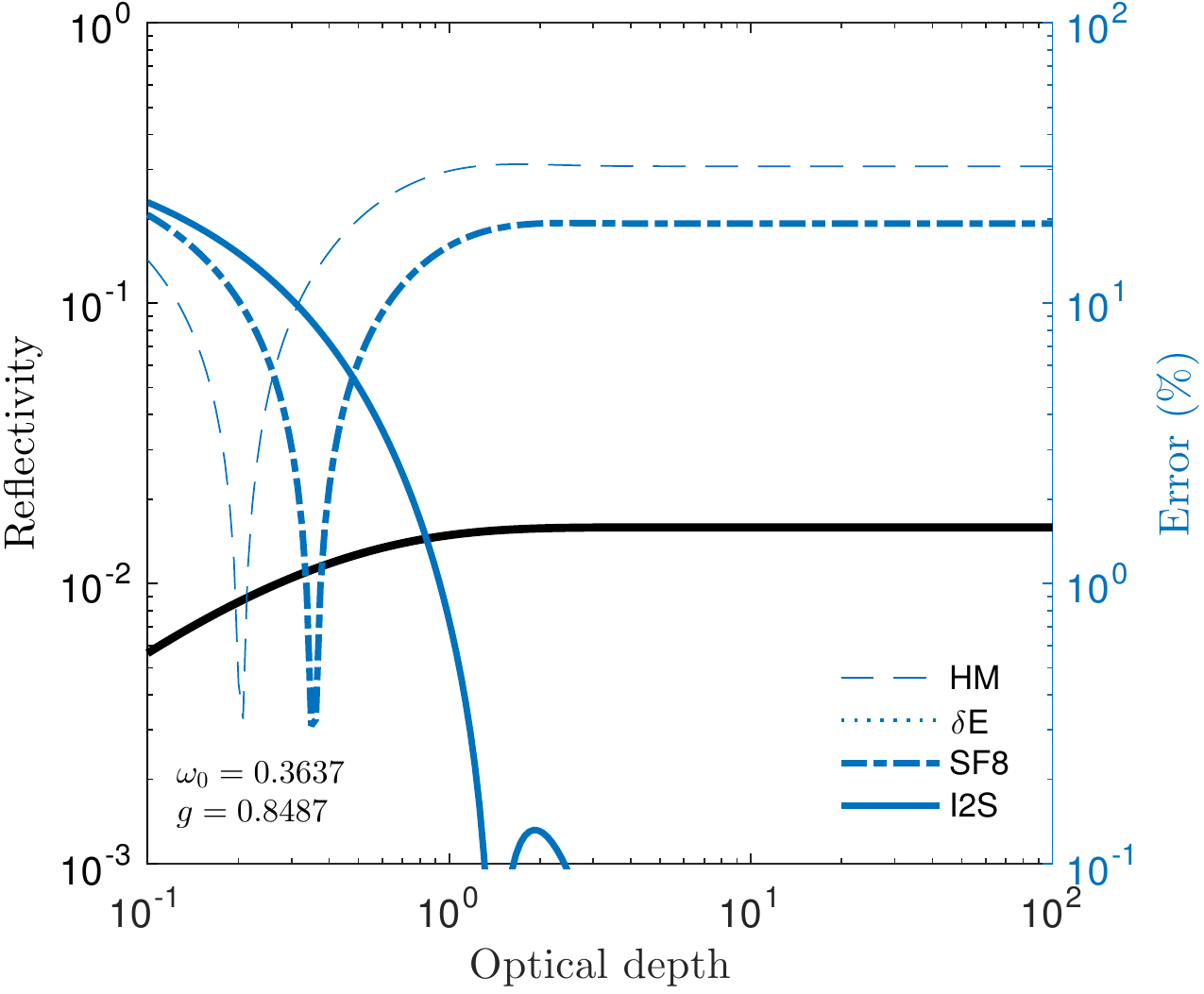}
\includegraphics[width=\columnwidth]{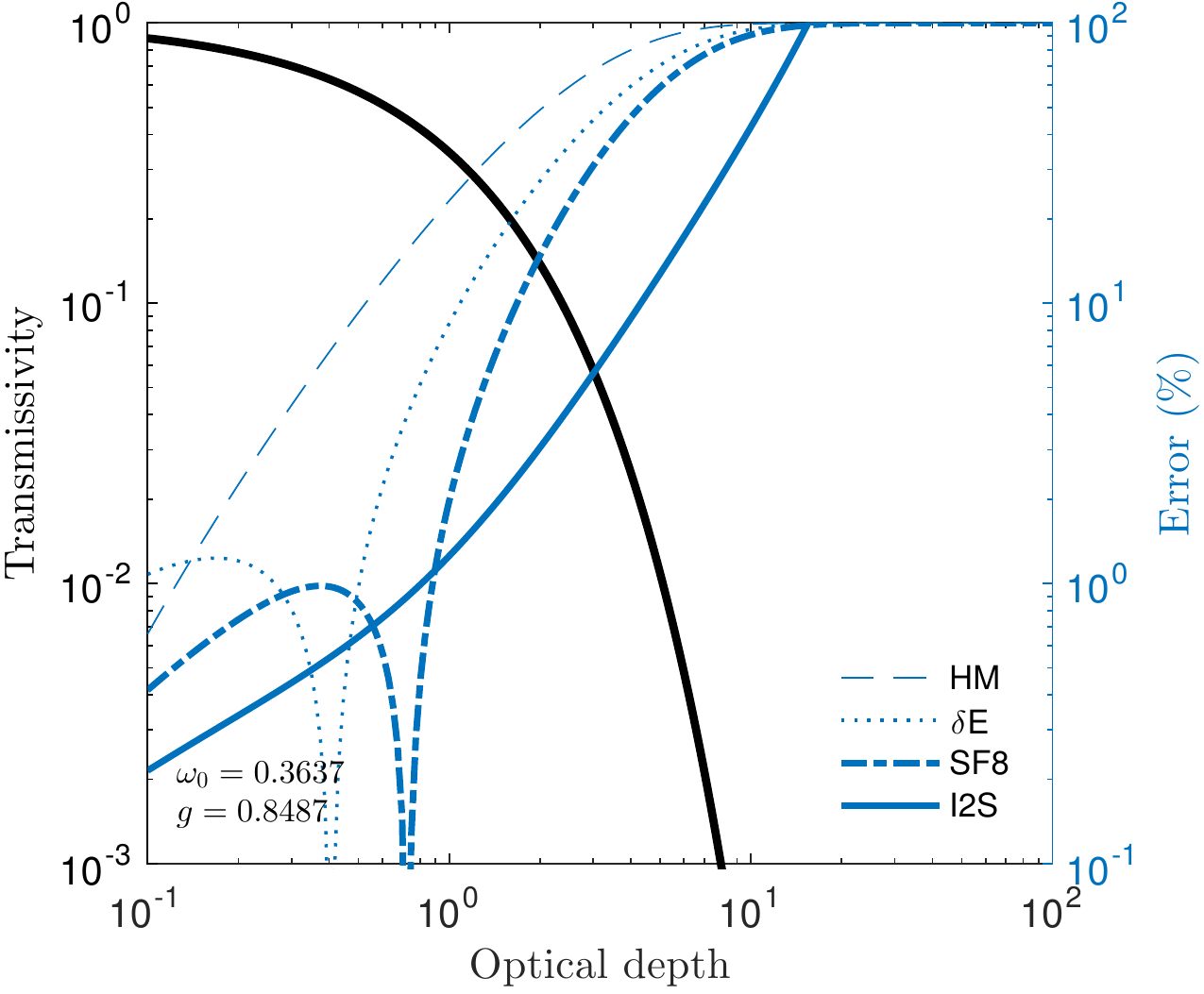}
\end{center}
\vspace{-0.2in}
\caption{Comparing calculations of the reflectivity (left column) and transmissivity (right column), which are the fractions of incident flux reflected by and transmitted through an atmospheric layer, respectively, as functions of the optical depth of the layer.  The labels ``HM", ``$\delta$E" and ``SF8" refer to the hemispheric two-stream, delta-Eddington and two-stream source function (with 8 Gaussian quadrature points in each hemisphere) methods.  Our improved two-stream method is denoted by ``I2S".  The reflectivities and transmissivities are shown as solid, black curves, while the percentage errors incurred by each method are shown as blue curves with various linestyles.  The three sets of values of $\omega_0$ and $g$ are chosen to facilitate comparison with \cite{toon89}.  For some combinations of $\omega_0$ and $g$, the $\delta$E method produces negative reflectivities \citep{mw80}, and we do not show them.}
\label{fig:compare}
\end{figure*}

\begin{figure}
\begin{center}
\includegraphics[width=0.9\columnwidth]{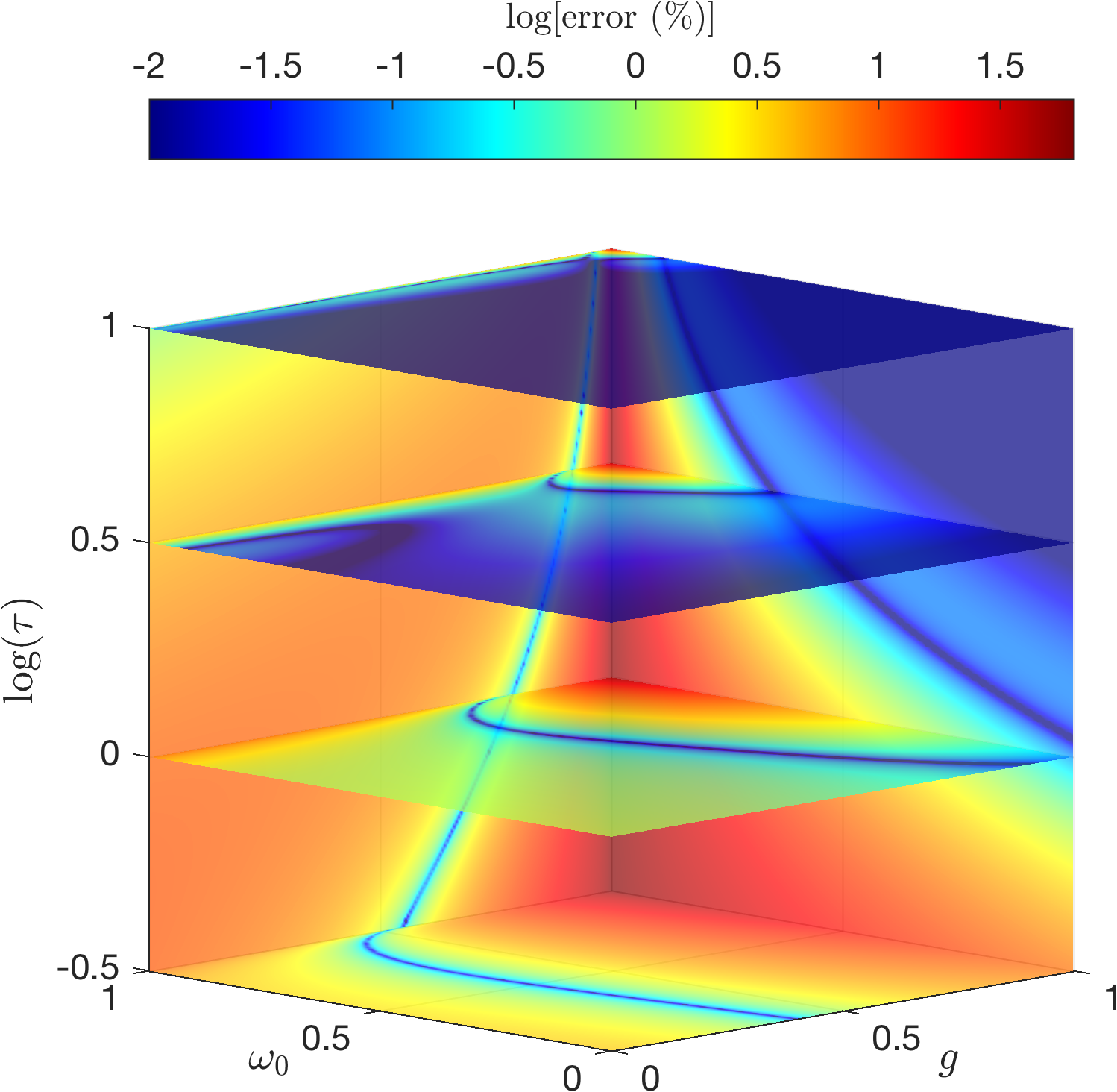}
\includegraphics[width=0.9\columnwidth]{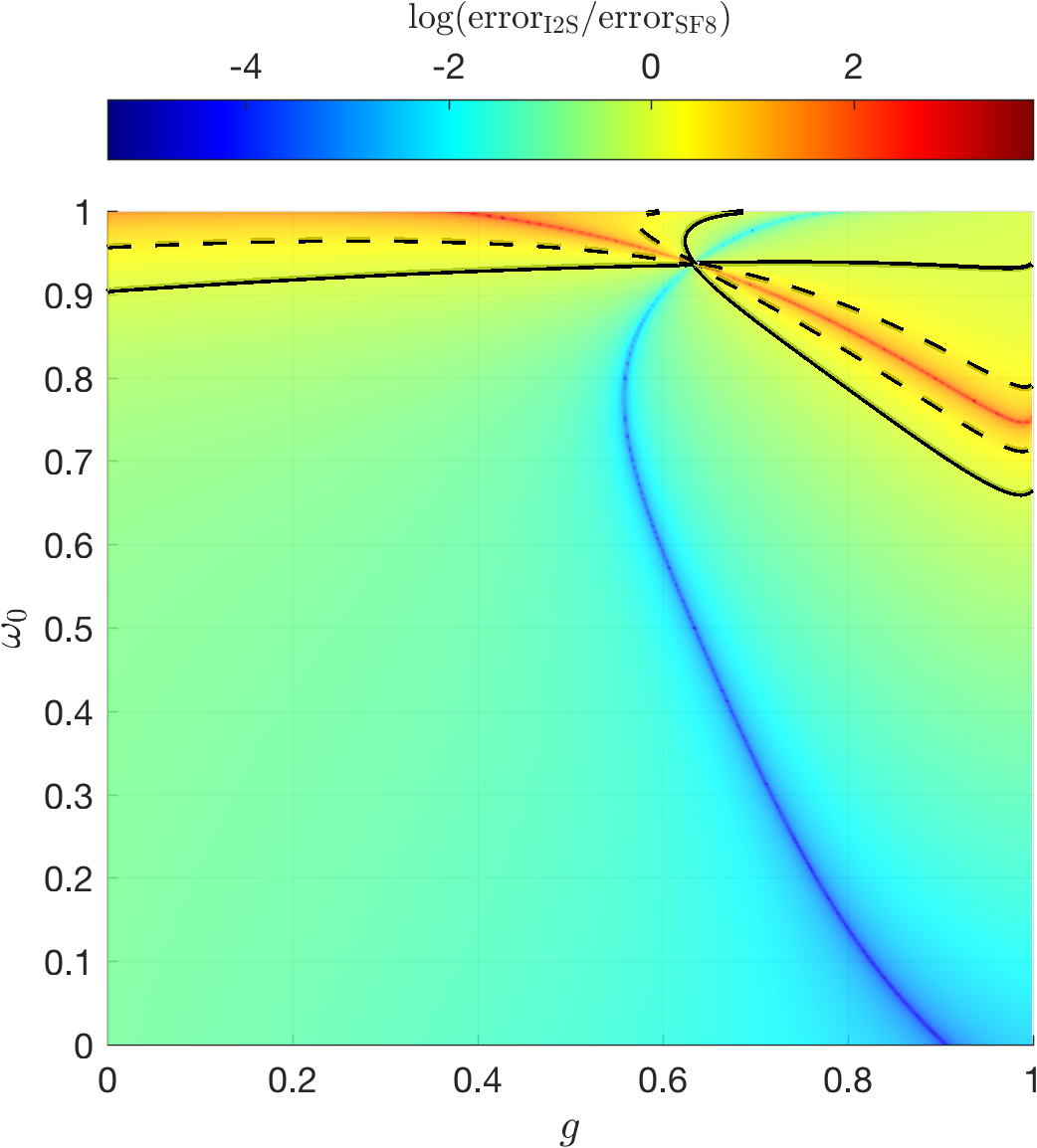}
\end{center}
\caption{Top panel: Error incurred by the improved two-stream method as a function of single-scattering albedo ($\omega_0$), scattering asymmetry factor ($g$) and optical depth ($\tau$) for calculations of the reflectivity.  Bottom panel: Ratio of errors of improved two-stream versus two-stream source function (with 8 Gaussian quadrature points in each hemisphere) methods for the reflectivity at $\tau=1$.  The solid black curves indicate regions of parameter space where the ratio of errors is exactly unity, while the dashed black curves correspond to a factor of 3.  The blue ``ridge" in the error plot where the improved two-stream calculation achieves an improvement of about 4 orders of magnitude, but this is because the reflectivity curves nearly intersect the reference \texttt{DISORT} curves.  Similarly, the red ``ridge" where the performance of the improved two-stream method becomes worse than a factor of 3, compared to the two-stream source function method, is due to the latter reflectivity curves nearly intersecting the reference \texttt{DISORT} curves.  These pathological situations occur when one multiplies or divides by a number that is very close to zero.}
\label{fig:error}
\end{figure}

We now compare calculations of the transmissivity and reflectivity to those performed using other commonly used methods: hemispheric two-stream \citep{hml14}\footnote{We specifically use the two-stream formalism written down by \cite{hml14}; we are \textit{not} claiming that \cite{hml14} should be solely cited for the two-stream method.}, delta-Eddington \citep{joseph76,wiscombe77,mw80} and two-stream source function \citep{toon89}.  The two-stream source function method is of particular interest, because it is widely implemented in the exo-atmospheres literature \citep{mm99,fortney08,cahoy10,morley13}.  It achieves a multi-stream solution using a clever mathematical trick: inserting the two-stream solution into the term of the radiative transfer equation involving the scattering phase function, which implies that the solution is, strictly speaking, not self-consistent.  We use the two-stream source function method with 8 streams in each hemisphere for a total of 16 streams, and sum up these streams numerically (weighted by the cosine of the polar angle) to construct the fluxes.

The delta-Eddington method uses the Eddington closure (see \citealt{pierrehumbert10} or \citealt{hml14}), but includes an additional feature: it approximates the scattering phase function as consisting of a Dirac-delta function and a series expansion involving the cosine of the scattering angle \citep{joseph76}.  The motivation behind this approximation is to attain higher accuracy for radiative transfer with large aerosols, which tend to produce a strong forward peak in the scattered intensity.  However, the delta-Eddington method has been criticized as being ad hoc, as the relative weighting of the Dirac-delta function and series terms is chosen arbitrarily \citep{wiscombe77}.\footnote{\cite{wiscombe77} describes how the delta-Eddington method includes the procedures of truncation and renormalization, describes them as being ``ad hoc" and remarks how ``none of the proposed variants is demonstrably superior to any other."  \cite{wiscombe77} further adds that ``the number of reasonable truncation and renormalization procedures is limited only by one's imagination."}  This criticism provided us with motivation to avoid the delta-Eddington method and its variants when constructing our improved two-stream method.  However, we include the delta-Eddington method in our comparisons because \textit{our emphasis is on reproducing Figures 2 and 3 of \cite{toon89} as a benchmarking exercise.}

We again use 32-stream \texttt{DISORT} calculations of the transmissivity and reflectivity as the ground truth.  Figure \ref{fig:compare} shows the transmissivity and reflectivity for three different sets of values of $\omega_0$ and $g$, chosen to faciliate comparison with the study of \cite{toon89}.  Most of the absorption or scattering of radiation by an atmospheric layer occurs at $\tau \sim 1$.  Dips in the error curves occur when the curve of the reflectivity or transmissivity intersects the \texttt{DISORT} curve, such that the error between them formally drops to zero.  In practice, it drops to \textit{nearly} zero, because the numerically-computed curves sampled at discrete points do not formally intersect.  The six sets of calculations in Figure \ref{fig:compare} suggest that our improved two-stream method achieves comparable or superior accuracy, compared to the other methods, often at the order-of-magnitude level, but with less computational effort.

Next, we focus on the error associated with the reflectivity, as it is known to be larger than for the transmissivity in atmospheric calculations involving aerosols \citep{kitzmann13}.  Figure \ref{fig:error} quantifies the error incurred by using our improved two-stream method as a function of $\omega_0$, $g$ and $\tau$.  When $\tau=10$, the error is $\sim 0.01\%$.  This is unsurprising, as it is by construction.  When $\tau=1$, the error is typically $\sim 1\%$, unless both $\omega_0 \approx 1$ and $g \approx 1$, in which case the error approaches 10\%.  In practice, the presence of gas, consisting of atoms and molecules, in the atmosphere reduces the value of $\omega_0$ to below unity because they provide a source of absorption (which increases the total cross section), meaning that these large errors are rarely encountered.  Figure \ref{fig:error} also shows the ratio of errors of the improved two-stream versus the two-stream source function methods.  For most combinations of $\omega_0$ and $g$, the improved two-stream method is more accurate than the two-stream source function method by an order of magnitude, unless $\omega_0 \gtrsim 0.9$.  This is remarkable, because it has the implementational simplicity of the two-stream method, but an accuracy that is superior to a 16-stream method.

\subsection{Toy model of early Martian atmosphere}

\begin{figure}
\begin{center}
\vspace{-0.2in}
\includegraphics[width=0.99\columnwidth]{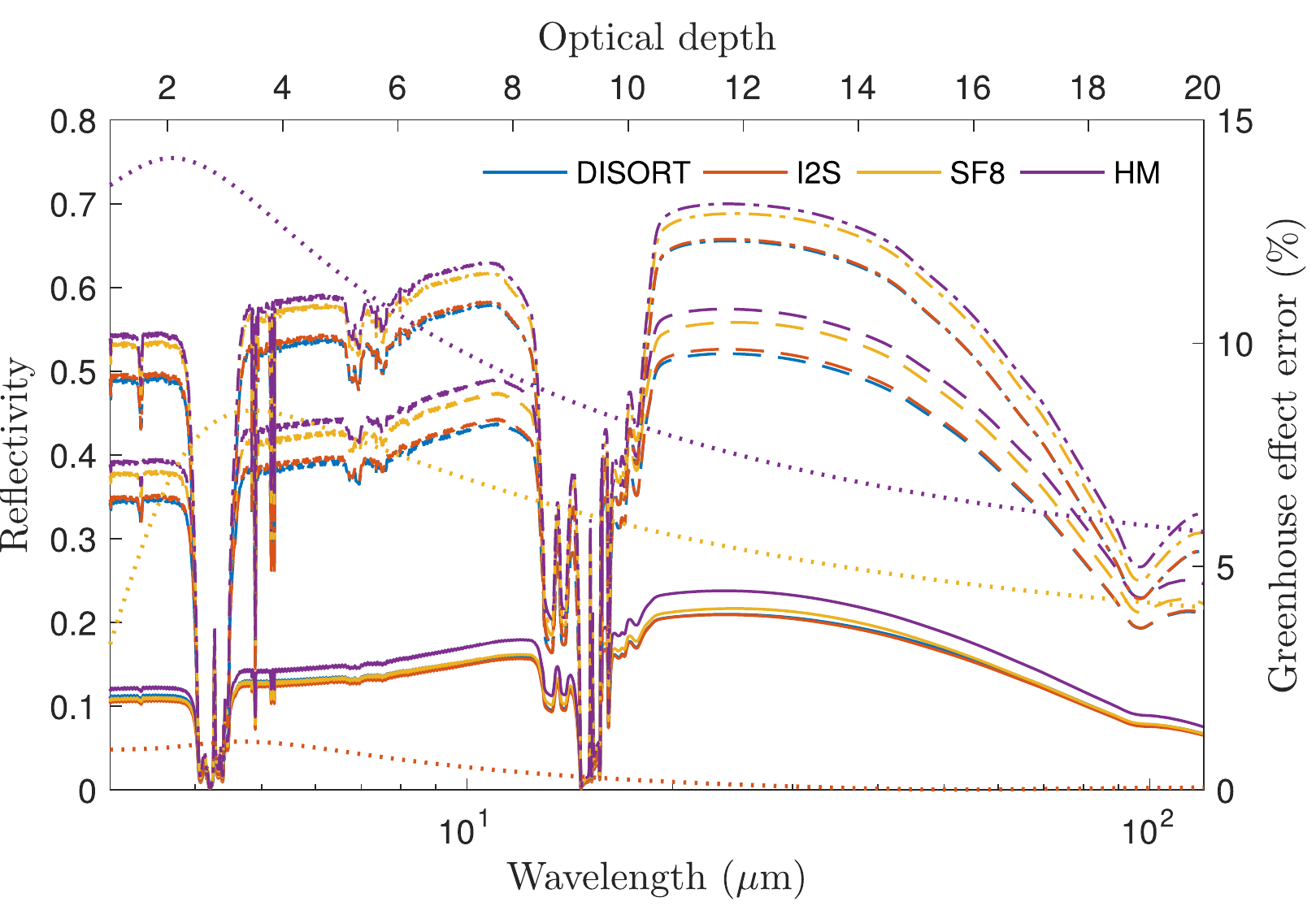}
\end{center}
\caption{Toy model of early Martian atmosphere populated by a cloud layer composed of carbon-dioxide ice particles.  The solid, dashed and dot-dashed curves are for $\tau=1, 5$ and 10, respectively.  The various colors refer to the method used.  The dotted curves are the wavelength-integrated errors, as functions of the optical depth, obtained by weighing the reflectivities by a Planck function with a temperature of 160 K, which is the condensation temperature of carbon dioxide at about 0.1 bar.  The labels ``HM", ``$\delta$E" and ``SF8" refer to the hemispheric two-stream, delta-Eddington and two-stream source function (with 8 Gaussian quadrature points in each hemisphere) methods.  Our improved two-stream method is denoted by ``I2S".}
\label{fig:ice}
\end{figure}

So far, we have examined calculations with fixed values of $\omega_0$ and $g$, because we were focused on benchmarking our improved two-stream method.  Real aerosols are associated with $\omega_0$ and $g$ that are \textit{functions} of wavelength \citep{draine03}.  To illustrate this behavior, we consider a toy model atmosphere of early Mars that hosts a cloud layer composed of medium-sized to large carbon-dioxide ice particles.  We do not compute a more realistic Mars model, because this has already been done in \cite{kitzmann16}.  Conceptually, understanding the interplay between aerosols and radiative transfer under Mars-like conditions is relevant to defining the outer edge of the classical habitable zone.

We imagine a scenario where starlight penetrates the atmosphere and heats up its surface (much like on Earth), which then re-emits the heat as infrared radiation.  The infrared radiation attempts to escape the atmosphere, but encounters the carbon-dioxide-ice cloud layer (assumed to be located at 0.1 bar), which reflects some of it back to the surface and heats up the atmosphere below the cloud layer.  The ice particles are assumed to follow a gamma distribution with an effective particle radius of 25 $\mu$m.  We assume that the atmosphere is dominated by gaseous carbon dioxide and that it has a temperature equal to the condensation temperature of carbon dioxide just below the cloud layer.  The absorptivity of the gas is parametrized by a grey opacity with a value chosen such that the optical depth contributed by the gas alone is 0.1.  We calculate the reflectivity of the cloud layer using realistic, wavelength-dependent single-scattering albedos and scattering asymmetry parameters (\citealt{kitzmann13} and Figure \ref{fig:properties}).  

In Figure \ref{fig:ice}, our improved two-stream calculations yield errors at 1\% or lower compared to 32-stream \texttt{DISORT} calculations.  By contrast, the hemispheric two-stream and two-stream source function methods yield errors at the 5--10\% level, depending on how opaque the cloud is.  As already eludicated by \cite{kitzmann16} using more realistic models of the early Martian atmosphere, these errors translate into an over-estimation of the surface temperature by about 40 K, enough to alter the qualitative conclusion.  Similarly, we anticipate that sophisticated simulations of the outer edge of the habitable zone using three-dimensional climate models would benefit from more accurate radiative transfer calculations using our improved two-stream method.

\section{Discussion}
\label{sect:discussion}

In the absence of scattering, the outgoing and incoming two-stream fluxes are decoupled.  The boundary condition at the bottom of the atmosphere (remnant heat from the formation of the exoplanet for gas giants and surface fluxes for rocky exoplanets) may be propagated upwards using the solution for the outgoing flux.  The boundary condition at the top of the atmosphere (stellar irradiation) may be propagated downwards using the solution for the incoming flux.  For each layer, one now has the values of the outgoing and incoming fluxes.  Taking the difference between these fluxes yields the net flux, which is then fed into the first law of thermodynamics to compute the change in temperature of each layer \citep{hml14}.  Updating the temperature in each layer in turn alters the opacities and the fluxes.  This iteration is performed numerically for a model atmosphere with multiple layers, until convergence is attained and radiative equilibrium is reached, e.g., see the implementation of \cite{malik17}.

When scattering is present, the pair of two-stream solutions feed into each other and an additional iteration is needed.  Physically, radiation may be scattered multiple times as it travels from a layer to its immediate neighbors and beyond.  It is worth emphasizing that this iteration is to enforce multiple scattering, and is distinct from the iteration for radiative equilibrium.  The alternative to such an iteration is to perform matrix inversion.  Instead of describing a pair of layers, the two-stream solutions may be used to describe the fluxes within a single layer bounded by two interfaces.  An atmosphere with a finite number of layers is represented by a set of equations, with each equation describing the fluxes at the layer center and interfaces.  Mathematically, this set of equations makes up a tridiagonal matrix, which may then be inverted using Thomas's algorithm \citep{arfken}.  Since Thomas's algorithm is essentially a set of recursive algebraic relations, the procedure remains efficient.  This property renders the solutions feasible for implementation in three-dimensional general circulation models (e.g., \citealt{showman09}), which require radiative transfer to be highly efficient in order to simulate climates for $\sim 10^7$ time steps or more.

\begin{acknowledgments}

We acknowledge financial support from the Swiss National Science Foundation, the PlanetS National Center of Competence in Research (NCCR), the Center for Space and Habitability (CSH) and the Swiss-based MERAC Foundation.  KH acknowledges a Visiting Professorship at Johns Hopkins University in the Zanvyl Krieger School of Arts and Sciences, held during the final revision and resubmission of this manuscript.

\end{acknowledgments}

\label{lastpage}

\end{document}